\documentclass[a4paper, 11pt]{article}
\usepackage[pdftex, a4paper]{geometry}
\usepackage{graphicx}				
				
\usepackage{amsmath}
\usepackage{amsfonts}
\usepackage{dsfont}
\usepackage{mathrsfs}
\usepackage{amssymb}
\usepackage{bbm}
\usepackage{epsfig}
\usepackage[english]{babel}
\usepackage[all]{xy}

\def\T{^{\rm\tiny T}}

\newtheorem{theorem}{Theorem}

\newtheorem{lemma}{Lemma}
\newtheorem{remark}{Remark}

\begin{document}

\title{\bf The Evolution of Network   Entropy in Classical and Quantum Consensus Dynamics}
\date{}

\author{Shuangshuang Fu, Guodong Shi,   Ian R. Petersen, and Matthew R. James\footnote{S. Fu, G. Shi, and M. R. James are with the Research School of Engineering, The Australian National University, Canberra, Australia. I. R. Petersen is with School of Engineering and Information Technology, University of New South Wales, Canberra, Australia. Email: shuangshuang.fu, guodong.shi, matthew.james@anu.edu.au,  i.r.petersen@gmail.com.}}
\maketitle

\begin{abstract}
In this paper, we investigate the evolution of the network entropy for consensus dynamics in classical or quantum networks. We show that in the classical case, the network entropy decreases at the consensus limit if the node initial values are i.i.d. Bernoulli random variables, and the network differential entropy is monotonically non-increasing if the node initial values are i.i.d. Gaussian. While for quantum consensus dynamics, the network's von  Neumann entropy is in contrast  non-decreasing. In light of this inconsistency, we  compare several  gossiping algorithms with random or deterministic  coefficients for classical or quantum networks, and show that quantum gossiping algorithms with deterministic coefficients are physically related to classical gossiping algorithms with random coefficients.
\end{abstract}

{\bf Keywords:} Consensus dynamics, Quantum networks, Entropy evolution

\section{Introduction}

With the basic idea being able to be  traced back to \cite{tsi}, problems of distributed consensus seeking  have been widely studied in the past decade  sparked  by the work of \cite{jad03,saber04}. The states of a group of interconnected nodes can asymptotically reach the average value of their initial states via neighboring node interactions and simple distributed control rule \cite{jad03,tsi,saber04}, which forms a foundational block for the further development in the broad range of control of network systems \cite{Magnus}.
The understanding of  distributed consensus algorithms has been substantially  advanced in aspects ranging  from convergence speed optimisation and directed links to  switching interactions and nonlinear dynamics \cite{xiao2005,ren05,Moreau,nedic09}.

On the other hand, recent work \cite{Ticozzi,Ticozzi-SIAM}  brought  the idea of distributed averaging consensus  to quantum networks, where each node corresponds to
a qubit, i.e., a quantum bit \cite{Nielsen}. In quantum mechanics, the state of a  qubit is represented by a density matrix over a two-dimensional Hilbert space $\mathcal{H}$, and the state of a quantum network with $N$ qubits corresponds to a density matrix over  the $N$'th tensor product of $\mathcal{H}$. The concepts regarding the network density matrix reaching a quantum consensus were systematically developed in \cite{Ticozzi}, and  it has been shown that a quantum consensus can be reached with the help of quantum swapping operators for both continuous-time and discrete-time dynamics \cite{Ticozzi,Ticozzi-MTNS,ShiDongPetersenJohansson}. In fact, the two categories of dynamics over classical and quantum networks can be put together into a group-theoretic framework \cite{Ticozzi-SIAM}, and quantum consensus dynamics can even be equivalently  mapped into some parallel classical dynamics over disjoint subsets of the entries of the network density matrix\cite{ShiDongPetersenJohansson}.

In this paper, we make an attempt to look at the relation between the two categories of dynamics from a  {\it physical} perspective, despite their various consistencies already shown in \cite{Ticozzi-SIAM,ShiDongPetersenJohansson}. The density matrix describes a quantum system in a mixed state that is   a statistical ensemble of several quantum states, analogous to the probability distribution function of a random variable \cite{Nielsen}. First of all we investigate the evolution of the network entropy for consensus dynamics in classical or quantum networks. We show that in  classical consensus dynamics, the network entropy decreases at the consensus limit if the node initial values are i.i.d. Bernoulli random variables, and the network differential entropy is monotonically non-increasing if the node initial values are i.i.d. Gaussian. While for the quantum consensus dynamics, the network's von  Neumann entropy is in contrast  non-decreasing. These observations suggest that the two types of consensus schemes may have different physical footings.  Then, we compare several  gossiping algorithms with random or deterministic  coefficients for classical or quantum networks and present novel convergence conditions for gossiping algorithms with random coefficients. The result shows that quantum gossiping algorithms with deterministic coefficients are physically consistent with classical gossiping algorithms with random coefficients.

The remainder of the paper is organized as follows. Section 2 presents the problem of interest as well as the main results. Section 3 presents the proofs of the statements. Finally Section 4 concludes the paper.

\section{Entropy Evolution and Classical/Quantum Gossiping}
 For a network with $N$ nodes in the set $\mathrm{V}=\{1,\dots,N\}$ with an interconnection structure given by the undirected graph $\mathrm{G}=(\mathrm{V}, \mathrm{E})$, the standard  distributed consensus control scheme is described   by the dynamics
\begin{align}\label{classical}
\frac{d}{dt} X(t) =-L_{\mathrm{G}} X(t),
\end{align}
where $X(t)=(X_1(t) \dots X_N(t))\T$ with $X_i(t)\in \mathds{R}$ representing  the state of node $i\in\mathrm{V}$, and $L_{\mathrm{G}}$ is the Laplacian of the graph $\mathrm{G}$. Here, we refer to \cite{Magnus} for a detailed introduction as well as for the definition of the graph Laplacian.

Also consider  a quantum network with $N$ qubits indexed  in the set $\mathrm{V}=\{1,\dots,N\}$. We can introduce a quantum interaction graph $\mathrm{G}=(\mathrm{V}, \mathrm{E})$, where $\{i,j\}\in\mathrm{E}$ specifies a swapping operator between the two qubits. The state of each qubit is represented by a density matrix over the two-dimensional Hilbert space $\mathcal{H}$, and the network state corresponds to a density matrix over $\mathcal{H}^{\otimes N}$, the $N$'th tensor product of $\mathcal{H}$.  Continuous-time quantum consensus control can be  defined by  \cite{Ticozzi-MTNS,ShiDongPetersenJohansson}
\begin{align}\label{quantum}
\frac{d}{dt}\rho(t)=\sum_{\{j,k\}\in \mathrm{E}}  \Big(U_{jk}\rho(t) U_{jk}^\dag  -\rho(t)\Big),
\end{align}
where $\rho(t)$ is the network density matrix, and $U_{jk}$ is the swapping operator between the qubits $j$ and $k$ (see \cite{Nielsen,Ticozzi} for details on the definition and realization of the swapping operators).

\subsection{Entropy Evolution}
Let the graph $\mathrm{G}$ be connected for either the classical or the quantum dynamics. It has been shown that for the system (\ref{classical}), there holds  (e.g., \cite{Magnus})
$$
X(\infty):=\lim_{t\to \infty} X(t)= \mathbf{1}\mathbf{1}\T X(0)/N,
$$
where $\mathbf{1}$ is the $N\times 1$ all-ones vector.  For the system (\ref{quantum}), there holds \cite{Ticozzi,Ticozzi-MTNS,ShiDongPetersenJohansson}
$$
\rho(\infty):=\lim_{t\to \infty} \rho(t)= \frac{1}{N!}\sum_{\pi \in \mathfrak{P}} U_\pi \rho(0) U_\pi^\dag,
$$
where $\mathfrak{P}$ is the permutation group and $U_\pi$ represents the quantum permutation operator induced by $\pi\in \mathfrak{P}$. The conceptual consistency of the systems (\ref{classical}) and (\ref{quantum}), as well as the logical consistency of the two  consensus limits, have been discussed in \cite{Ticozzi,ShiDongPetersenJohansson}.

The Shannon entropy is a fundamental  measure of uncertainty of a random variable \cite{CoverBook}. The entropy $H(Z)$, of a discrete random variable $Z$ with  alphabet $\mathcal{Z}$ is defined as
$$
H(Z):=-\sum_{z\in \mathcal{Z}} p(z)\log p(z).
$$
Here  $\log$ is to the base $2$ and $p(\cdot)$ is the probability mass function. The differential entropy $h(Z)$ of a continuous random variable $Z$ with density $f(z)$ is defined as $$
h(Z):=-\int_{\mathcal{S}} f(z) \log f(z)dz,
$$
where $\mathcal{S}$ is the support of $Z$.  As a natural generalization of the Shannon entropy, for a quantum-mechanical system described by a density matrix $\rho$, the von Neumann entropy is defined as \cite{Nielsen}
$$
S(\rho)=-{\rm tr}(\rho \log \rho),
$$
where  ${\rm tr}(\cdot)$ is the trace operator.

We present the following result  for classical consensus dynamics.

\begin{theorem}\label{thmclassical} (i) Let $X_i(0)$ be independent and identically distributed (i.i.d.) Bernoulli random variables with mean $p\in (0,1)$. Then $NX(\infty)$ obeys binomial distribution. Therefore,   for the system (\ref{classical}), there holds $H(X(0))=N\big[p\log p^{-1}+(1-p)
\log (1-p)^{-1} \big]$, and
$
H(X(\infty))\simeq \frac{1}{2} \log \big( 2\pi eNp(1-p)\big)+ O(\frac{1}{N}).$

(ii) Let $X_i(0)$ be i.i.d. Gaussian random variables with mean $\mu$ and variance $\sigma^2$. Then $h(X(t))$ is a non-increasing function over $[0,\infty)$ for the system (\ref{classical}).
\end{theorem}

Here $H(X(t))$ and $h(X(t))$ are defined for the random vector $X(t)$. For the Gaussian case,   $h(X(\infty))$ does not yield a finite number  since $X(\infty)=\mathbf{1}\mathbf{1}\T X(0)/N$ becomes degenerate. We can however  conveniently use $h(X(\infty))=h(X_i(\infty))$ and a simple  calculation gives
$$
h(X(0))=\frac{N}{2}\big[\log (2\pi e\sigma^2)\big],\ \ \ h(X(\infty))=\frac{1}{2}\log \big( \frac{ 2\pi e\sigma^2}{N}\big).
$$

 For quantum consensus dynamics, the following result holds.

\begin{theorem}\label{thmquantum}
For the system (\ref{quantum}), $S(\rho(t))$ is a non-decreasing function over $[0,\infty)$.
\end{theorem}

The above results reveal that, the network entropy in general decreases with classical consensus dynamics, but increases with quantum consensus dynamics. This appears to be surprising  noticing their consistencies pointed out in \cite{Ticozzi,ShiDongPetersenJohansson}. However,    although the systems (\ref{classical}) and (\ref{quantum}) can be formally united (cf., \cite{ShiDongPetersenJohansson}), $X(t)$  represents a random variable in the classical world, while $\rho(t)$ is a probability mass function by its definition.

\subsection{Numerical Examples}
\begin{figure}[t]
\begin{center}
\includegraphics[height=1.8in]{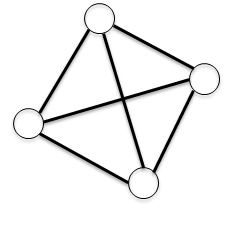}
\caption{The quantum interaction graph $\mathsf{G}$ with $4$ (classical or quantum) nodes.}
\label{Fig1}
\end{center}
\end{figure}

We now provide a simple example illustrating the derived results. Consider a graph with $4$ (classical or quantum) nodes as shown in Figure \ref{Fig1}.

\begin{figure}
\begin{minipage}[t]{0.5\linewidth}
\centering
\includegraphics[width=3.2in]{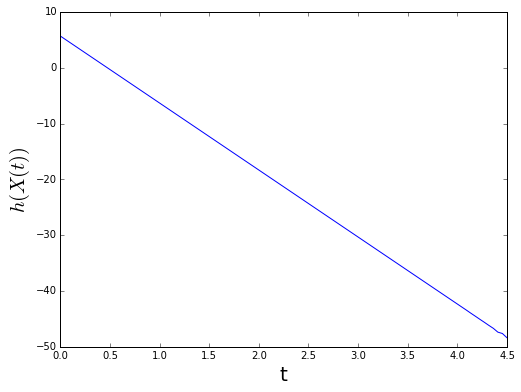}
\end{minipage}%
\begin{minipage}[t]{0.5\linewidth}
\centering
\includegraphics[width=3.2in]{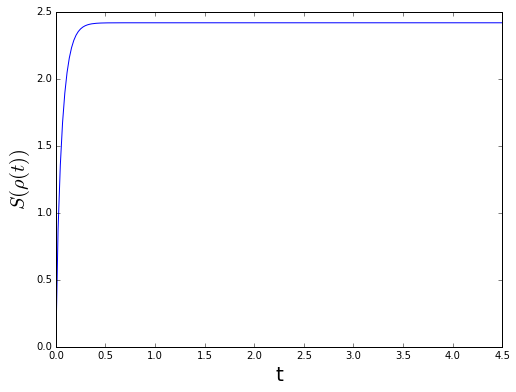}
\end{minipage}
\caption{The evolution of the network entropy for classical (left) and quantum (right) dynamics, respectively. }
\label{fig:entropy}
\end{figure}

For the classical case, we take the $X_i(0)$ as an i.i.d. standard Gaussian random variable. For the quantum case, we take the initial density matrix as
$$
\rho(0)=|01+-\rangle\langle 01+-|
$$
under the Dirac notion \cite{Nielsen}.
The evolution of the differential entropy and the von  Neumann entropy with the classical and quantum consensus dynamics is plotted, respectively, in Figure \ref{fig:entropy}.

\subsection{Gossiping with Random/Deterministic  Coefficients}

In this subsection, we provide a {\it physical} perspective to explain  the observations in Theorem~\ref{thmclassical} and Theorem~\ref{thmquantum} by investigating a serial of classical or quantum gossiping algorithms with random or deterministic coefficients.

A random gossiping process is defined as follows. Consider $N$ nodes in the set $\mathrm{V}$ with an underlying interaction graph $\mathrm{G}$ which is undirected and connected.  Time is sequenced by $k=0,1,\dots$.    At time $k$, a node $i$ is first drawn with probability $1/N$, and then node $i$ selects another  node $j$ who shares a link with node $i$ in the graph $\mathrm{G}$ with probability $1/{\rm deg}(i)$. Here ${\rm deg}(i)$ is the degree of node $i$ in the graph $\mathrm{V}$. In this way, a random pair $\{i,j\}$ is selected.  Additionally, let $b_k, k=0,1,\dots$ be a sequence of i.i.d. Bernoulli random variables with mean $1/2$, which are also independent of any other possible  randomness.

\begin{itemize}
\item
In the classical case, each node $i$ holds a real-valued state $X_i(k)\in \mathds{R}$ at time $k$. Their initial states,  $X_1(0),\dots,X_N(0)$, are assumed to be $N$ (not necessarily independent) random variables over a common underlying probability space. The marginal probability (mass or density) distribution of node $X_i(k)$ is denoted as $p^i_k(\cdot)$.  When the pair  $\{i,j\}$ is selected at time $k$, only the two selected nodes update their values and we consider the following  algorithms.

\begin{itemize}
\item[{[A1]}]  {\it (Classical Gossiping with Deterministic Coefficients, \cite{Boyd-gossip})} Node $i$ and $j$  update their values as
\begin{align}\label{random}
X_i(k+1)=X_j(k+1)=\frac{1}{2} X_i(k)+\frac{1}{2} X_j(k).
\end{align}
\item [{[A2]}] {\it (Classical Gossiping with Random Swapping, \cite{Shi-CDC})} Node $i$ and $j$  update their values as
\begin{equation}
\begin{split}
&X_i(k+1)=b_kX_i(k)+(1-b_k) X_j(k);\\
&X_j(k+1)=(1-b_k) X_i(k)+b_kX_j(k).
\end{split}
\end{equation}
\end{itemize}

\item In the quantum case, each node $i$ represents a qubit and    $\rho(k)$ is the network density matrix at time $k$. When the qubit pair  $\{i,j\}$ is selected at time $k$, we correspondingly consider the following algorithms.

\begin{itemize}
\item[{[AQ1]}]  {\it (Quantum Gossiping with Deterministic Coefficients, \cite{Ticozzi})} The quantum network updates its density matrix as
\begin{align}\label{random}
\rho(k+1)=\frac{1}{2} \rho(k)+\frac{1}{2} U_{ij} \rho(k)U_{ij}^\dag.
\end{align}
\item [{[AQ2]}] {\it (Quantum Gossiping with Random Swapping, \cite{Ticozzi-SIAM})} Node $i$ and $j$  update their values as
\begin{equation}
\rho(k+1)=b_k \rho(k)+(1-b_k)U_{ij}\rho(k)U_{ij}^\dag.
\end{equation}
\end{itemize}
\end{itemize}

We state a few immediate  facts for the algorithms [A1], [A2], [AQ1], and [AQ2].
\begin{itemize}
\item[(i)] The evolution of $\mathbf{E}\{X(k)\}$ is exactly the same along with the algorithms [A1] and [A2]. Similar conclusion holds also for the algorithms [AQ1] and [AQ2].
\item[(ii)] Algorithms [A1] and [AQ1] are {\it algorithmically equivalent}, in the sense that [AQ1] can be divided into a set of parallel algorithms in the form of [A1] over disjoint entries of $\rho(t)$ (see \cite{ShiDongPetersenJohansson} for a thorough treatment via vectorizing $\rho(t)$). Similarly, the algorithms [A2] and [AQ2] are algorithmically equivalent.
\item[(iii)] Algorithms [A2] and [AQ1] are {\it physically equivalent}, in the sense that for a sequence of  underlying random variables $X(k)$ evolving along [A2], their joint probability mass/density function, denoted  $f_k(x_1,\dots, x_N)$ (which is exactly the physical interpretation of the density matrix $\rho(k)$) will  evolve in the form of   [AQ1] (cf., \cite{Ticozzi-SIAM}):
    \begin{align}
    f_{k+1}(x_1,\dots, x_N)=\frac{1}{2}f_k(x_1,\dots,x_i,\dots,x_j,\dots, x_N)+\frac{1}{2}f_k(x_1,\dots,x_j,\dots,x_i,\dots, x_N)
    \end{align}
    if the pair $\{i,j\}$ is selected at time $k$.
\end{itemize}

Recall that a Markov chain is ergodic if it is both aperiodic and irreducible \cite{Durr}. We present the following result establishing the limiting behaviors of the algorithm [A2], which is consistent with the observations of the entropy evolution in Theorems \ref{thmclassical} and \ref{thmquantum} as well as the point (iii) above.
\begin{theorem}\label{thmfooting}
For the  algorithm [A2], there holds that
\begin{itemize}
\item [(i)]  $\big\{X(k)\big\}_{k=0}^\infty$ forms an ergodic Markov chain given $X(0)$;

\item [(ii)] $\lim_{k \rightarrow \infty}p_k^i(\cdot)={\sum_{i=1}^N p^i_0(\cdot)}/{N}$, where  the convergence is exponentially fast under the distance induced by $\ell^1$ (for $X(0)$ given by discrete random variables) or  $\mathcal{L}^1$ (for continuous $X(0)$) norms.
\end{itemize}
\end{theorem}

\begin{remark}
One can also consider the case in a gossiping process when two selected node $i$ and $j$  update their values by (Classical Gossiping with Random Coefficients)
\begin{align}\label{random}
{\rm [A1']} \quad \quad X_i(k+1)=X_j(k+1)=b_kX_i(k)+(1-b_k) X_j(k).
\end{align}
From the second  Borel-Cantelli Lemma (e.g., Theorem 2.3.6. in \cite{Durr}), that  almost surely, $X_i(k)$ reaches  a common value for all $i\in \mathrm{V}$ in finite time along the algorithm [A1']. Interestingly, it is easy to see that the evolution of the $p^i_k(\cdot)$ is the same along the algorithms [A1'] and [A2].
\end{remark}

\begin{remark}
The scheme of the algorithms [A2] was briefly discussed in Section 6.2 of \cite{Ticozzi-SIAM}, which is also a form of gossiping algorithms with unreliable but perfectly dependent link communications studied in \cite{Shi-CDC} with mixing coefficient one. Here Theorem \ref{thmfooting} advances the previous understandings by showing that the algorithm [A2] defines an ergodic Markov chain for any given initial condition as well as presenting the detailed convergence properties of the marginal distribution functions for both discrete and continuous $X(0)$.
\end{remark}

\begin{remark}
We assume that the mean of the $b_k$ is $1/2$ just for the ease of presentation. It is clear from the proof that Theorem \ref{thmfooting} holds for arbitrary $\mathbf{E}\{b_k\}\in(0,1)$. The ergodicity plays an essential role in the convergence of the marginal distributions: the case with $\mathbf{E}\{b_k\}=0$ fails because $X(k)$ is no longer aperiodic; the case with $\mathbf{E}\{b_k\}=1$ fails because $X(k)$ is no longer irreducible.
\end{remark}

\section{Proofs of Statements}
This section provides the proofs of Theorems \ref{thmclassical}, \ref{thmquantum} and \ref{thmfooting}.
\subsection{Proof of Theorem \ref{thmclassical}}

\noindent (i) The fact that $H(X(0))=N\big[p\log p^{-1}+(1-p)
\log (1-p)^{-1} \big]$ follows straightforwardly from the independence of the $X_i(0)$. On the other hand, $X_i(\infty)=\sum_{j=1}^N X_j(0)/N$ follows a Binomial distribution $\mathcal{B}(N, p)$ whose entropy is well-known to be $\frac{1}{2} \log \big( 2\pi eNp(1-p)\big)+ O(\frac{1}{N})$. Since $X_i(\infty)=X_j(\infty) $ for all $i,j\in\mathrm{V}$,
there holds $H(X(\infty))=H(X_i(\infty))$. This proves (i).

\medskip

\noindent (ii) The solution $X(t)$ of the system (\ref{classical}) is
\begin{align}
X(t)=e^{-tL_{\mathrm{G}} } X(0).
\end{align}
As a result, for any $t\geq 0$, $X(t)$ is a Gaussian random vector. Then
\begin{align}\label{1}
h(X(t))&=\frac{1}{2}\log\Big |(2\pi e)^N \mathbf{E} \Big[[X(t)-\mathbf{E}(X(t))][X(t)-\mathbf{E}(X(t))]\T \Big]\Big|\nonumber\\
&= \frac{1}{2}\log\Big |(2\pi e \sigma^2)^N e^{-2tL_{\mathrm{G}} } \Big|,
\end{align}
where $|\cdot|$ represents the matrix determinant.

We take $\epsilon>0$ and compare $h(X(t+\epsilon))$ with $h(X(t))$. There holds  from (\ref{1}) that
\begin{align}
h(X(t+\epsilon))= h(X(t))+ \frac{1}{2}\log\big | e^{-2 \epsilon L_{\mathrm{G}} } \big|.
\end{align}
Since $L_{\mathrm{G}}$ is the Laplacian of a connected undirected graph $\mathrm{G}$, $L_{\mathrm{G}}$ has a unique zero eigenvalue, and all non-zero eigenvalues of $L_{\mathrm{G}}$ are positive  (cf., \cite{Magnus}).  Consequently, all eigenvalues of  $e^{-2 \epsilon L_{\mathrm{G}}}$ are positive and no larger than one, which yields that
$$
\big| e^{-2 \epsilon L_{\mathrm{G}} } \big|\leq 1.
$$
This proves $h(X(t+\epsilon))\leq h(X(t))$. Since $\epsilon$ is chosen arbitrarily, we conclude that  $h(X(t))$ is a non-increasing function. The calculations of $h(X(0))$ and $h(X(\infty))$ are straightforward.

We have now completed the proof of  Theorem \ref{thmclassical}.  \hfill$\square$

\subsection{Proof of Theorem \ref{thmquantum}}

The proof relies on the following lemma.

\begin{lemma}\label{lem1}
Let $\epsilon>0$ and fix $s\geq 0$. For the system (\ref{quantum}), there exist $m_{\pi}(\epsilon)\geq 0, \pi\in \mathfrak{P}$  with $\sum_{\pi \in \mathfrak{P}} m_{\pi}(\epsilon)= 1$ such that
\begin{align}
\rho(s+\epsilon)=\sum_{\pi \in\mathfrak{P}} m_{\pi}(\epsilon) U_{\pi}\rho(s) U_{\pi}^\dag.
\end{align}
\end{lemma}
{\it Proof.} Define a set
$$
\Sigma_s= {\rm co}\Big( U_{\pi}\rho(s) U_{\pi}^\dag: \pi\in \mathfrak{P}\Big),
$$
where ${\rm co}(\cdot)$ stands for the  convex hull. It is straightforward to see that $U_{jk}\rho U_{jk}^\dag\in \Sigma_s$ if    $\rho\in \Sigma_s$. As a result, $\Sigma_s$ is an invariant set of the system (\ref{quantum}) in the sense that $\rho(t)\in \Sigma_s$ for all $t\geq 0$ as long as $\rho(0)\in\Sigma_s$. The desired lemma thus follows immediately. \hfill$\square$

Recalling that the von Neumann entropy $S(\rho)$ is a concave function of $\rho$, and that $S(\rho)=S(U\rho U^\dag)$ for any unitary operator $U$, we conclude from Lemma \ref{lem1} that \begin{align}
S(\rho(s+\epsilon))&= S\Big( \sum_{\pi \in\mathfrak{P}} m_{\pi}(\epsilon) U_{\pi}\rho(s) U_{\pi}^\dag\Big)\nonumber\\
&\geq \sum_{\pi \in\mathfrak{P}} m_{\pi}(\epsilon)S\Big( U_{\pi}\rho(s) U_{\pi}^\dag\Big)\nonumber\\
&=S(\rho(s))
\end{align}
for any $\epsilon>0$ and $s\geq 0$ in light of the fact that $U_\pi$ is unitary for all $\pi\in \mathfrak{P}$. This proves that $S(\rho(t))$ is a non-decreasing function and Theorem \ref{thmquantum} holds. \hfill$\square$

\subsection{Proof of Theorem \ref{thmfooting}}
\noindent (i). First of all it is clear that $\big\{X(k)\big\}_{k=0}^\infty$ is  Markovian  from its definition. Recall that $\mathfrak{P}$ is the $N$'th permutation group. We denote the permutation matrix associated with   $\pi\in\mathfrak{P}$ as $M_\pi$. In particular, the permutation matrix associated with the swapping between $i$ and $j$ is denoted as $M_{\pi_{ij}}$. The state transition of  $\big\{X(k)\big\}_{k=0}^\infty$ along the algorithm A2 can be written as
\begin{align}
\mathbf{P}\Big(M_{\pi_{ij}} X(k)  \Big | X(k)\Big)= \big(1/{\rm deg}(i)+1/{\rm deg}(j)\big)/N,\ \ \{i,j\}\in\mathrm{E}.
\end{align}
Since the graph $\mathrm{G}$ is connected, the swapping permutations defined along the edges of  $\mathrm{G}$ form a generating set of the permutation group $\mathfrak{P}$. Consequently, given $X(0)$, the set
$$
\Big\{M_\pi X(0), \ \pi\in\mathfrak{P}\Big\}
$$
is the state space of $X(k)$, which contains  at most $N!$ elements. Finally it is straightforward to verify that for any given $X(0)$,  $X(k)$ is irreducible and aperiodic, and therefore forms an ergodic Markov chain.

\medskip

\noindent (ii). The statement is in fact a direct consequence from the ergodicity of $X(k)$. We however need to be a bit more careful since we assume that $X(0)$ takes value from an arbitrary (not necessarily discrete) probability space and the $X_i(0)$ are not necessarily independent. We denote the state transition matrix for $X(k)$ as $P\in \mathds{R}^{N\times N}$. We calculate $p^i_k(\cdot)$ from basic probability equality $\mathbf{P}(A)=\sum_{s=1}^m \mathbf{P}(A|C_i)$ under  $\sum_{i=1}^N \mathbf{P}(C_i)=1$ and $\mathbf{P}(C_i \bigcap C_j)=0$, and then immediately obtain
\begin{align}
p^i_k(\cdot)= \sum_{s=1}^N e_s\T P^k e_i p^s_0(\cdot),
\end{align}
where $e_i$ is the unit vector with the $i$'th entry being one. It is clear that the above calculation does not rely on $X(0)$  being discrete or continuous, and $p^i_k(\cdot)$ represents probability mass or density function wherever appropriate. From the definition of the algorithm A2 $P$ is a symmetric matrix and the ergodicity of $X(k)$ leads to
\begin{align}
\lim_{k\to \infty}P^k ={\mathbf{1}\mathbf{1}\T}/{N}
\end{align}
at an exponential rate. The desired conclusion thus follows.
\section{Conclusions}
We have  investigated the evolution of the network entropy for consensus dynamics in classical or quantum networks. In the classical case, the network entropy decreases at the consensus limit if the node initial values are i.i.d. Bernoulli random variables, and the network differential entropy is monotonically non-increasing if the node initial values are i.i.d. Gaussian. For quantum consensus dynamics, the network's von  Neumann entropy is on the contrary non-decreasing. This observation can be easily generalized to  balanced directed graphs \cite{saber04,Ticozzi-MTNS,Quantum-Directed}. In light of this inconsistency, we also compared several  gossiping algorithms with random or deterministic  coefficients for classical or quantum networks, and showed that quantum gossiping algorithms with deterministic coefficients are physically consistent with classical gossiping algorithms with random coefficients.

\end{document}